\documentclass[a4paper,12pt]{article}
\usepackage{graphicx}
\usepackage[utf8]{inputenc}
\usepackage{latexsym}
\usepackage{amssymb}
\usepackage{amsmath}
\usepackage{graphics}
\usepackage{float}
\begin{document}
	
	\title{Odderon: models {\it vs} experimental data - a short review of recent papers}
	
	\author{Evgenij Martynov  \thanks{Bogolyubov Institute for Theoretical Physics,
			Metrolologichna 14b, Kiev, UA-03143, Ukraine} \thanks{email: martynov@bitp.kiev.ua} \and
		Basarab Nicolescu \thanks{Faculty of European Studies, Babes-Bolyai University, Emmanuel de Martonne Street 1, 400090 Cluj-Napoca, Romania} \thanks{email: basarab.nicolescu@gmail.com}}
	\maketitle

\begin{abstract}
The surprising low TOTEM datum $\rho_{pp}$ = 0.09$\pm$0.01 at 13 TeV \cite{TOTEM-1} generated an important flux of papers, which can be classified in three categories: 1) papers which claim that this result is the first experimental discovery of the Odderon, namely in its maximal form; 2) papers which tried, without success, to find alternative explanations of this result; 3) papers which contest the discovery of the Maximal Odderon.

In the present short note we discuss two recent papers belonging to the third category.
\end{abstract}

O.V. Selyugin and J.R. Cudell \cite{Selyugin-Cudell} make the claim that the contribution of the Maximal Odderon is small at $t=0$ and $\sqrt{s} $ = 13 TeV. This claim is wrong for the following reasons:
\begin{enumerate} 
\item 
The authors of Ref.~\cite{Selyugin-Cudell}   include a non-Maximal Odderon in their HEGS model and impose that $\rho_{pp}$ is approximately equal to 0.12, namely $\rho_{pp}$ = 0.13$\pm$0.015 at 13 TeV. This value is obviously far from the TOTEM experimental value. A lowering of 0.03-0.04 of the $\rho$-value is impossible to obtain without the Maximal Odderon, which therefore is required by the measured $\rho$ value.
\item
The same authors get a value of $\sigma_t$  = 106$\pm$2.5 mb at 13 TeV and they claim that this result is in agreement with the TOTEM value $\sigma_t$  = 110.6$\pm$3 mb. However, in getting their result, they allow an arbitrary additional normalization coefficient which, they say, "reflects the systematic errors``. In fact, this way of treating systematic errors is in contradiction with the experimental systematic errors. Such an arbitrary lowering of $\sigma_t$ TOTEM value (corresponding to a behavior of $\sigma_t$ milder than $\ln^2s $) is not allowed.
\item 
The same authors consider in their fits only the data above $\sqrt{s}$=100 GeV, because they say that they want "to avoid possible problems connected with the low-energy region``. This procedure is not satisfactory because it is precisely in the region below 100 GeV, namely at 52.8 Gev, a quite strong signal on the Odderon was obtained at ISR \cite{ISR-52-8}.
\item 
On a more theoretical level, the authors claim that the real parts of the final scattering amplitude grow asymptotically like $\ln  s$ "as required by the analytical properties of the amplitude`` (non-Maximal-Odderon), and not like $\ln^2s $ (Maximal Odderon). This statement is wrong as proved in the past, in the first paper introducing the Maximal Odderon \cite{LN}. 
The claim of the authors is based upon a result of J. Finkelstein  {\it et al} \cite{Finkelstein et al}, which is by no means a rigorous theorem but an elaborated model in the framework of eikonal unitarization of the Odderon as a single pole in complex momentum plane. Moreover, the authors themselves of Ref.~\cite{LN} assert this fact, not excluding that in future one can find a way to accommodate in the respective approach the Maximal Odderon \cite{Chung-I Tan}.
\end{enumerate}

\medskip

L. Harland-Lang, V. Khoze, A. Martin and M. Ryskin (HLKMR) \cite{HLKMR} go further and they strongly claim that the Odderon is "elusive``, in other words is not seen at LHC. In fact the Durham group V. Khoze, A. Martin and M. Ryskin  (KMR) published in the last year a series of contradictory preprint versions: in some of these papers the Odderon is needed and in others it is not. This simply shows that the "global`` KRM model is not stable in its predictions. 
But before showing why the claim of an "elusive Odderon`` is not correct, let us clarify an important theoretical point. 

KMR repeatedly say that the Maximal Odderon violates unitarity. This statement is wrong and has its source in a logical misunderstanding. We clarified this point more than a quarter of century ago. In 1992 in a rigorous way, a class of amplitudes with the Maximal Odderon type of asymptotics and simultaneously consistent with s-channel unitarity, fixed-t analyticity and the absence of  $J=1$ massless state in the $t$-channel was considered \cite{GLN}. We explicitly stated that, in our case, the complex phase in the impact parameter amplitudes is bounded when s is going to infinity, in contradistinction to the case of models yielding black disk asymptotics. But KMR are using precisely the black disk asymptotics. Therefore they make a logical vicious circle: they eliminate from the beginning the Maximal Odderon and, at the end, they deduce that the Maximal Odderon violates unitarity. Let us note in passing that KMR never quoted Ref.~\cite{GLN}.

The unitarization method to solve the Finkelstein-Kajantie problem which is considered by KRM 
is  correct only at Black Disk Limit. Moreover, their solution  does not work in the case of multigap diffractive production processes because of Abramovsky-Gribov-Kancheli cancelations (exactly as for one particle inclusive production in the central region of rapidity). Generally, if  KRM $S^2(b)$-method does not work it does not mean that unitarity is violated. It means that another method of unitarity recovery should be applied. The simplest method based on the Dyson-Schwinger equations for Froissaron propagators and 3F-vertexes (depending on angular momenta of Froissarons) was suggested in Ref.~\cite{Ball}. It can be extended for FMO model. 

Now we show why the claim of an "elusive Odderon`` is not correct:
\begin{enumerate}
\item 
The form of the Odderon took by HLKMR (see Figure on $\rho$ in Ref.~\cite{HLKMR}) is oversimplified and has nothing to do neither with the Maximal Odderon nor with the general theory of the Odderon.
\item 
They are in disagreement with the UA4 $\bar pp$ data on $d\sigma/dt$ and TOTEM 7 TeV $pp$ data (see Figure on $d\sigma/dt$  in Ref.~\cite{HLKMR}).
\item They are in complete disagreement with the $d\sigma/dt$ D0 $\bar pp$ data, crucial for showing explicit Odderon effects when took in comparison with the $d\sigma/dt$ TOTEM data at 2.76 TeV \cite{TOTEM-2} (which are not quoted by HLKMR).

\begin{minipage}[][][l]{6cm}
{\item 
	Generally, the effects of the Maximal Odderon were not observed at the experiments at lower than LHC energy (excepting the observation of the difference in $pp$ and $\bar pp$ in the region of the dip at $\sqrt{s}$=52.8 GeV \cite{ISR-52-8}) because the real part of Maximal Odderon becomes visible only at energies higher than about 2 TeV. It can be seen from the Figure 1, where the partial contributions are  shown. They are calculated using the parameters obtained in the Ref.~\cite{MN-1}.}
\end{minipage}
\hspace {0.5 cm}
\begin{minipage}[][][l]{7.5 cm}
	\includegraphics[width=0.85\linewidth]{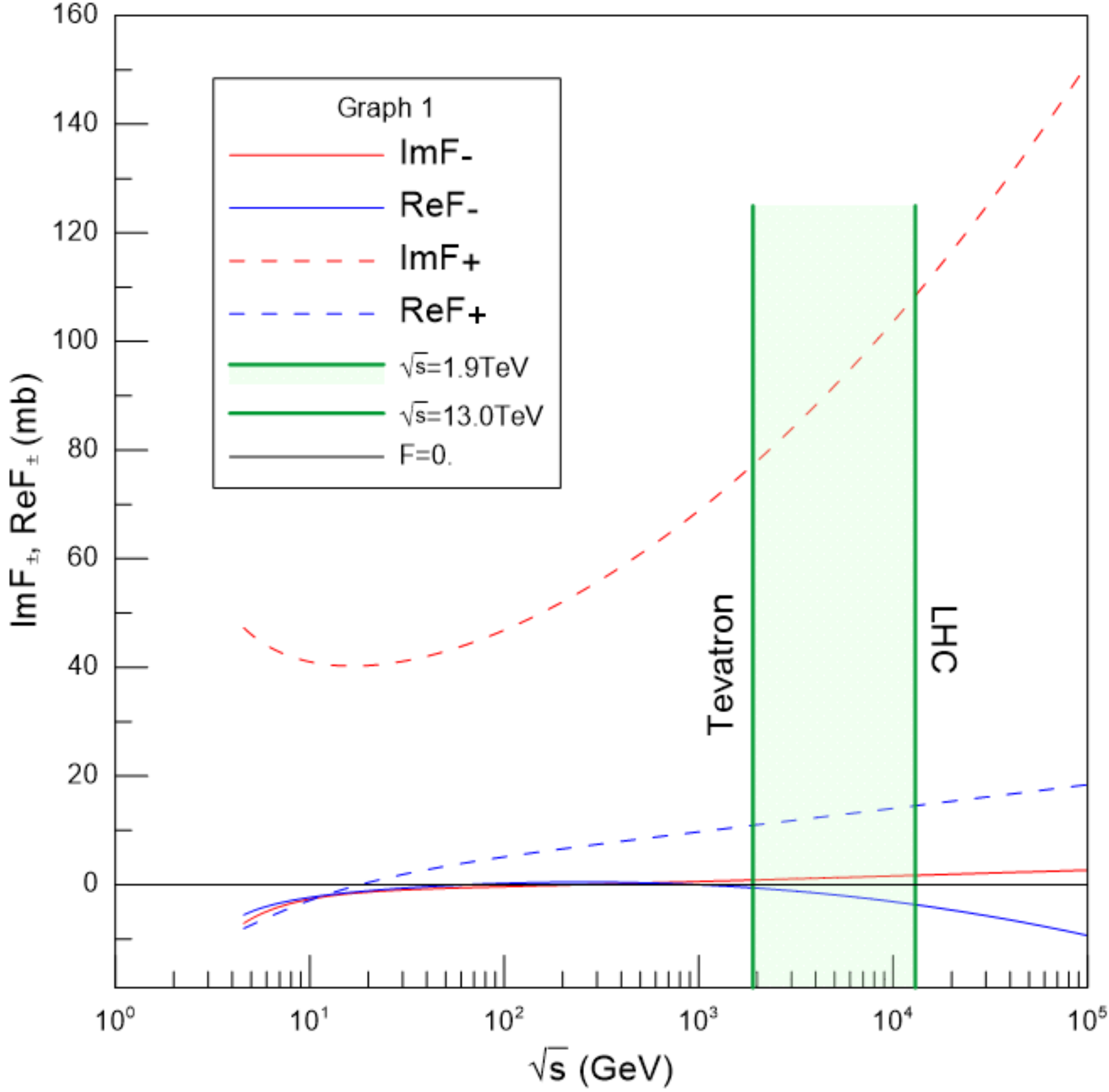}
	
		The solid blue (red) line is the real (imaginary) part of the crossing-odd contribution (Eqs. (9, 11, 13) from \cite{MN-1})
\end{minipage}
\item 
HLKMR completely neglect the important model-independent result of Cs\"{o}rg\"{o} {\it et  al.}~\cite{Csorgo}, which is in total agreement with the Froissaron-Maximal Odderon model \cite{FMO}: the slope in $pp$ scattering has a different behavior in t than the slope in $\bar pp$ scattering. This phenomenon is a clear Odderon effect.
\end{enumerate}

In conclusion, models which try to negate the Odderon effects violate the $pp$ TOTEM data, the $\bar pp$ D0 data or both on measured observables such as the total cross-section, $\rho$, the dip-region in the differential cross-section. Actually, in 2018 the existence of Odderon is fully established by TOTEM pp data at the LHC, both at $t=0$ and $t$ different from 0 (as well as by D0 $\bar pp$ data at the Tevatron). Ultimately, the Odderon was discovered experimentally.


\begin{thebibliography}{}
	\bibitem{TOTEM-1} G. Antchev  {\it et al.}, CERN-EP-2017-335, 2017.
	\bibitem{Selyugin-Cudell} O.V. Selyugin and J.R. Cudell, arXiv:1810.11538 [hep-ph]; talk by O.V. Selyugin at the Conference "Diffraction and Low-x 2018``, 26 August 2018 - 1 September 2018, 
	Reggio Calabria, Italy 
	\bibitem{ISR-52-8} A. Breakstone {\it et  al.}, Phys. Rev. Lett. 54 (1985) 2180; S. Erhan, {\it et al.}, Phys. Lett.  B152 (1985) 131.
	\bibitem{LN}  L. {\L}ukazsuk, B. Nicolescu, Lett. al Nuovo Cimento {\bf 8} (1973), 405.
	\bibitem{Finkelstein et al} J. Finkelstein,  H.M. Freid, K. Kang and C-I Tan, Phys. Lett. 232 (1989) 257.
    \bibitem{Chung-I Tan} Chung-I Tan, private communication, XLVIII International Symposium on Multiparticle Dynamics (ISMD2018),  Singapore, 3-7 September, 2018. 
	\bibitem{GLN} P. Gauron,  L. {\L}ukazsuk, B. Nicolescu, Physics Letters, {\bf B 294} (1992) 298.
	\bibitem{HLKMR} Lucian Harland-Lang, Valery Khoze, Alan Martin, Misha Ryskin, Talk by A. Martin at the 6th Workshop on QCD and Diffraction joint with Various Faces of QCD,   Krakow,   November 15-17,   2018.
	\bibitem{Ball} James S. Ball, Nucl. Phys. B102 (1976) 347.
	\bibitem{TOTEM-2}Talk of  Simone Giani on behalf of the TOTEM collaboration, LHCC RRB meeting, 30 October 2018, https://indico.cern.ch/event/757043/.
	\bibitem{MN-1} E. Martynov, B. Nicolescu,  Phys. Lett., {\bf B778} (2018) 414.
    \bibitem{Csorgo}  T. Cs\"{o}rg\"{o}, R. Pasechnik and A. Ster, arXiv: 1807.02897 [hep-ph].
	\bibitem{FMO} E. Martynov, B. Nicolescu, arXiv:1808.08580v2 [hep-ph].
\end{thebibliography}
\end{document}